\documentclass[amsmath,twocolumn,superscriptaddress, showkeys]{revtex4-2}

\usepackage[utf8]{inputenc}
\usepackage{graphicx}
\usepackage{dcolumn}
\usepackage{bm}
\usepackage{textcomp}  
\usepackage{siunitx}
\usepackage{braket}
\usepackage{commath} 
\usepackage{parskip}
\usepackage{graphicx}
\usepackage{xcolor}
\usepackage{amsfonts}
\usepackage{tabularx}
\usepackage{array}
\usepackage{lipsum}  

\usepackage{wrapfig}

\begin{document}
\title{Fluorescence-detected two-dimensional electronic spectroscopy of a single molecule}

\author{Sanchayeeta Jana}
\affiliation{Experimental Physics III, University of Bayreuth, 95447 Bayreuth, Germany}

\author{Simon Durst}
\affiliation{Experimental Physics III, University of Bayreuth, 95447 Bayreuth, Germany}

\author{Markus Lippitz}
\email{markus.lippitz@uni-bayreuth.de}
\affiliation{Experimental Physics III, University of Bayreuth, 95447 Bayreuth, Germany}

\date{\today}



\begin{abstract}
\begin{wrapfigure}[16]{r}{82.5mm}
\hspace*{-30mm}\includegraphics[width=82.5mm]{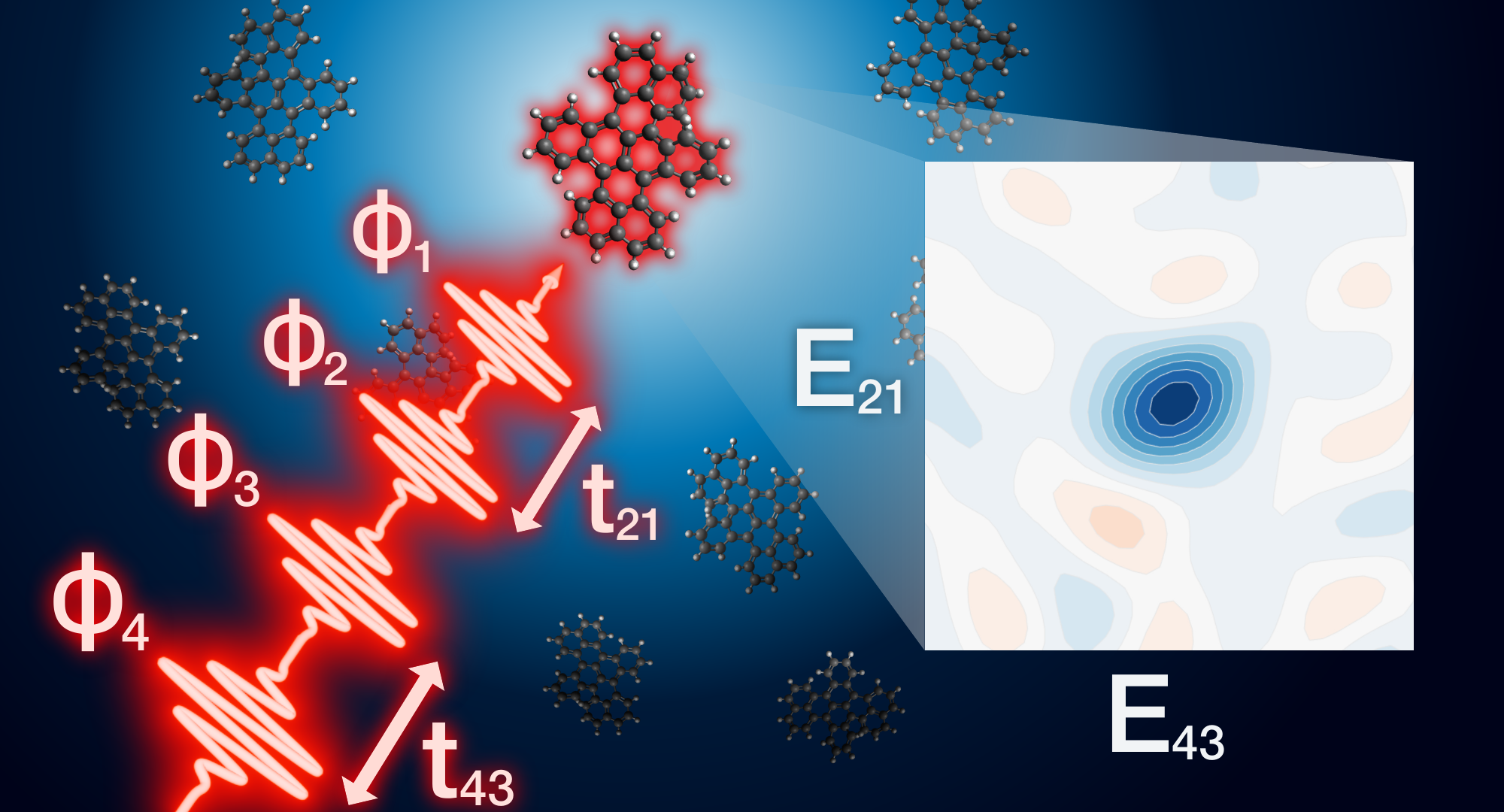}
\end{wrapfigure}
Single-molecule fluorescence spectroscopy is a powerful method that avoids ensemble averaging, but its temporal resolution is limited by the fluorescence lifetime to nanoseconds at most.
At the ensemble level, two-dimensional spectroscopy provides insight into ultrafast femtosecond processes such as energy transfer and line broadening, even beyond the Fourier limit, by correlating pump and probe spectra.
Here, we combine these two techniques and demonstrate 2D spectroscopy of individual molecules at room temperature using the example of dibenzoterrylene (DBT) in a polymer matrix. We excite the molecule in a confocal microscope with a phase-modulated train of femtosecond pulses and detect the emitted fluorescence with single-photon counting detectors. Using a phase-sensitive detection scheme, we were able to measure the nonlinear 2D spectra of most of the DBT molecules we studied. 
Our method is applicable to a wide range of single emitters and opens new avenues for understanding energy transfer in single quantum objects on ultrafast time scales.
\end{abstract}

\maketitle

Single-molecule fluorescence imaging and spectroscopy is a well-established technique with a wide range of applications \cite{Adhikari2022}. It avoids ensemble averaging of spectroscopic properties and provides access to their temporal evolution without the need to synchronize the ensemble \cite{moerner2002}. However, the accessible time scales are limited by the fluorescence lifetime. Faster processes, such as coherence decay and excitation transport, are hidden from the observer. Nonlinear optical spectroscopy is required to access these ultrashort time scales, i.e., a single emitter must interact with two photons within a short time delay.

The simultaneous interaction of a single dye molecule with two photons is a challenging task. Only a handful of nonlinear experiments have been successfully performed on a single molecule at room temperature: The conceptually simpler ones use two laser pulses, similar to pump-probe spectroscopy \cite{EvanDijk05,Brinks2010,Hildner2013a,Weigel15,Maly2016,Wilma19,Schedlbauer2020,Moya2022}. However, due to the Fourier limit, high temporal resolution is linked to low spectral resolution. Fourier transform spectroscopy by a delayed pulse pair preserves the temporal resolution of the short laser pulses and obtains the spectral resolution from the Fourier transform. This approach was applied to single molecules by Liebel~et~al.~\cite{Liebel2018}, replacing the second ('probe') pulse by a pulse pair, and by Fersch~et~al.~\cite{fersch2023}, replacing the first ('pump') pulse. In both cases, the other pulse remained spectrally broad.

In the most versatile form, both pump and probe are phase-stabilized pulse pairs, allowing to  correlate pump and probe spectra. This technique is called two-dimensional electronic spectroscopy \cite{tanimura1993, hamm1998, Gelzinis2019,Maiuri2020}. Traditionally, coherence radiation is detected by interference with a local oscillator. However, the incoherent action of the excitation can also be used, for example,  fluorescence emission \cite{Tekavec2007}.  These two approaches show strong similarities but are not completely identical \cite{kuhn2020, Maly2020,Bolzonello2023}.  Fluorescence-detected two-dimensional spectroscopy has been used to study small ensembles of light-harvesting complexes \cite{Karki2019, Tiwari2018} and a nanostructured film of zinc phthalocyanine \cite{Goetz2018}. 

In this work, we demonstrate the ultimate sensitivity of fluorescence-detected two-dimensional spectroscopy by investigating a single dye molecule at a time. We show that approximately $10^6$ detected photons are sufficient to obtain useful nonlinear spectra, which is only a moderate requirement for nano-spectroscopy at room temperature.

\begin{figure*} 
    \includegraphics*[]{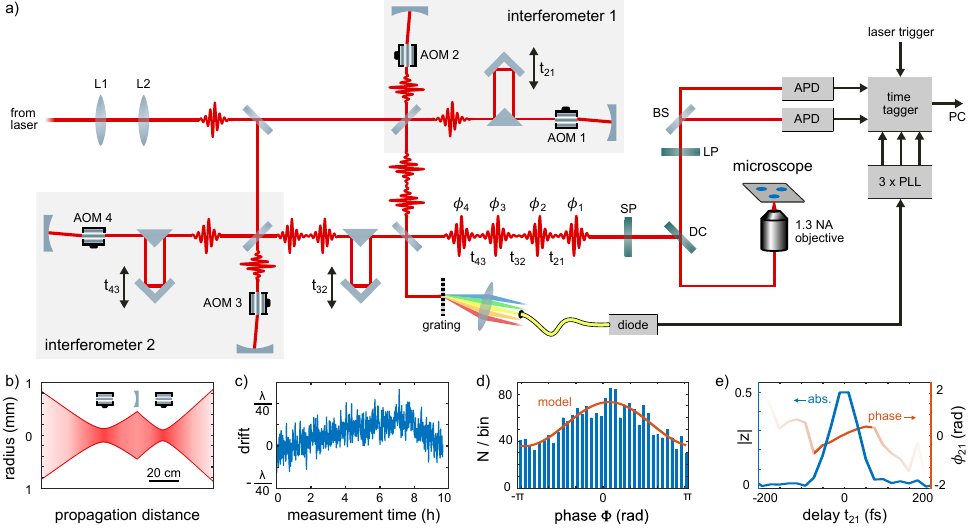}
    \caption{\label{fig:setup} \textbf{Experimental setup for single-molecule 2D spectroscopy} a)~A cascaded interferometer generates four pulses with controllable interpulse delays $t_{ij}$. Acousto-optic modulators (AOM) tag each field with a phase $\phi_i$. The pulse sequence excites a single molecule in a microscope. Fluorescence photons from the molecule are detected by two single-photon counting detectors (APD). A time tagger registers the photons and the reference phases via phase-locked loops (PLL). b)~A spherical mirror folds back the Gaussian beam to allow spectrally broad operation of the AOMs.  c)~Long-term stability of the interferometer.   d)~Phase distribution of photons at zero delay. e) Demodulated linear lock-in signal as a function of interpulse delay. }
\end{figure*}

\textbf{Experimental setup} 
We use the rapid phase-cycling scheme \cite{Tekavec2007}. An input laser pulse is split into four delayed parts in a sequence of interferometers (Fig.~\ref{fig:setup}a and Supplementary Material, section \ref{sm:setup}). Each field $E_i$ is 'tagged' by slightly shifting its optical frequency by $\Omega_i$ in an acousto-optic modulator, thus rapidly modulating its phase $\phi_i(t) = \Omega_i t$. The interactions of the light fields with the molecule carry these phases. Different linear and nonlinear processes that lead to a population of the excited state and thus to fluorescence can be distinguished by their dependence on the phases $\phi_i$ and cause an amplitude modulation of the fluorescence intensity by a characteristic combination of the modulation frequencies $\Omega_i$. In the experiment, we chose $\Omega_i \approx 2 \times 78$~MHz with differences $\Omega_{ij} = \Omega_i - \Omega_j \approx 10 \dots 50$~kHz. One-photon excitation by the fields $i$ and $j$ is thus modulated with the frequency $\Omega_{ij}$, two-photon excitation by these fields with $2 \Omega_{ij}$. A Fourier transform along the delay $t_{ij}$ between these two pulsed fields leads to the corresponding excitation spectra.

In two-dimensional spectroscopy, four fields interact with the molecule, leading to the rephasing and non-rephasing nonlinear signals. In the rephasing process, the coherent state during the delay $t_{43}$ is the complex conjugate of the state during the delay $t_{21}$, while it is the same for the non-rephasing signals. The former is similar to a photon echo experiment. In rapid phase cycling, we detect the non-rephasing signal at the modulation frequency $\Omega_{21} + \Omega_{43}$ and the rephasing signal at $\Omega_{21} - \Omega_{43}$.

As proposed by Tekavec et al. \cite{Tekavec2007}, we derive the frequencies $\Omega_{ij}$ not directly from the function generator driving the modulators but from a spectrally filtered optical output of the interferometers. This has two advantages: we follow all vibrations of the interferometer, which also phase shift the optical fields. This eliminates the need for external stabilization. In addition, the spectral filter defines a reference wavelength that places the detection in a rotating frame. Consequently, all spectra are measured relative to this reference wavelength, causing the interferometer fringes to oscillate much slower with the delays, reducing the number of measurement points required. In our case, a grating combined with a single mode fiber results in a reference spectral width of about 1.5~nm. This ultimately determines our spectral resolution. By comparing the phase of the reference signal to the phase of the electronic drive, we measure the path length difference between the interferometer arms. We use this information to linearize our delay stages. Fig.~\ref{fig:setup}c shows such a signal over 9~hours, demonstrating the passive stability of the interferometers.

We differ from Refs. \cite{Tekavec2007, Tiwari2018} by using the acousto-optic modulators in double-pass to minimize their spectral dispersion. Since the ultrasonic wave in the modulator acts as a grating, the direction of propagation after the modulator is different for each spectral component. A spherical mirror folds back all components provided that the modulator to mirror distance is equal to the radius of curvature of the mirror. For optimal alignment, the modulator should be located at the waist of a Gaussian beam (Fig.~\ref{fig:setup}b), and the distance to the mirror should be greater than the Rayleigh range so that the wavefront coincides with the mirror surface.

The four pulses then excite the molecules in a home-built inverted microscope consisting of a high NA (1.3) objective and a piezo stage. The fluorescence photons emitted by the molecules are collected by the same objective and detected by two single-photon counting detectors. A time tagger records the arrival time $t_n$ of each photon. Zero crossings ($T_m$) of the rapid phase cycling at three frequencies $\Omega_{21}$, $\Omega_{32}$, and $\Omega_{43}$ are determined by phase-locked loops and also recorded by the time tagger. In the offline analysis of this data stream, we generate for each excitation process of interest the corresponding mixing phase $\Phi(t)$ from the zero crossings $T_m$.  We can display the photons as a histogram over this modulation phase $\Phi$ and find a cosine-shaped modulation (Fig.~\ref{fig:setup}d). We determine the complex modulation amplitude $z$ of the photon stream by summing over all $N$ photons in this pixel 
\begin{equation}
    z = \frac{2}{N} \, \sum_{j = 1}^N \cos \Phi(t_j) - i \sin \Phi(t_j) \quad .
\end{equation}
The prefactor is chosen so that $|z| \le 1$.  This is lock-in detection of discrete events \cite{uhl2021}: The term under the sum is the mixer, and we implemented a synchronous filter by choosing $N$ such that 
\begin{equation}
    \Phi(t_N) \le  \Phi(t_1) + m \, 2\pi  <  \Phi(t_{N+1}) 
\end{equation}
with the maximum possible integer $m$, i.e. we sum over full $2 \pi$ periods in $\Phi$. As a function of the interpulse delay, the modulation amplitude $z$ gives the interferogram in the rotating frame (Fig.~\ref{fig:setup}e). The Fourier transform along the delay $t_{21}$ gives the 1D excitation spectrum $Z(\omega_{21})$, and similarly for $t_{32}$ and $t_{43}$.  The phase spectrum $\phi(\omega_{ij})$ must be zero at the frequency of the rotating frame since the linear excitation spectrum is real and positive \cite{Agathangelou2021}. This allows us to determine the phase offset of the setup and correct the zero delay position using the first derivative of the phase spectrum.

\begin{figure} 
    \includegraphics[]{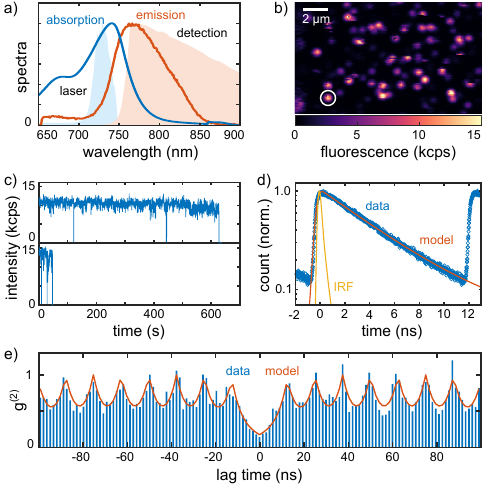}
    \caption{\textbf{Single molecules of dibenzoterylene (DBT)} a)~Absorption and emission spectra of DBT together with our spectral excitation and detection window. 
    b)~Confocal fluorescence imaging of single immobilized DBT molecules. c)~Fluorescence time trace of two exemplary molecules. d)~Time-correlated single photon counting (TCSPC) of a single DBT molecule e) Cross-correlation $g^{(2)}$  fitted with a single emitter model. The anti-bunching dip at $\tau = 0$ indicates a single quantum emitter.}
    \label{fig:DBT_molecule}
\end{figure}

\begin{figure*}
    \includegraphics*[]{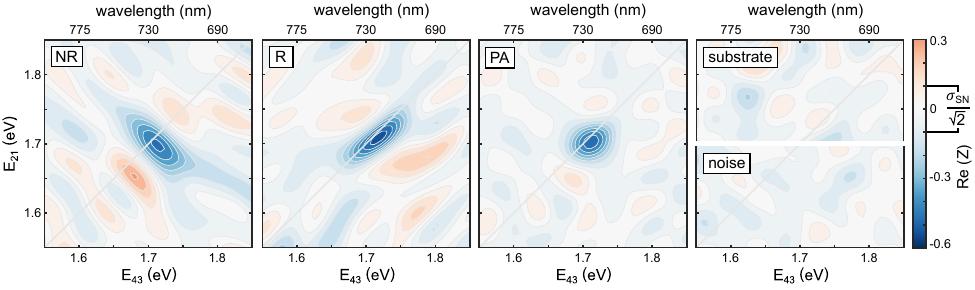}
    \caption{\textbf{2D spectra of single DBT molecule} The real part of non-rephasing (NR), rephasing (R), and purely absorptive (PA) 2D spectra of a single DBT molecule.
    Reflected laser light (substrate) of the same intensity only contains noise that agrees with the amplitude of simulated shot-noise traces (noise).
}
    \label{fig:2D_spectra}
\end{figure*}

\textbf{Sample} We chose dibenzoterylene (DBT) molecules in a polymethyl methacrylate (PMMA) matrix as the sample to demonstrate our technique (see also Supplementary Material, section \ref{sm:sample}). DBT is a well-known dye in single molecule experiments \cite{toninelli2010, Erker2022}. Our laser system is a Ti:Sa oscillator (80 MHz). Its spectrum only partially overlaps with the absorption spectrum of the dye (Fig.~\ref{fig:DBT_molecule}a). Using a pulse shaper, we achieve a pulse length of about 40 fs on the sample, slightly longer than the expected Fourier limit of 28 fs. We find individual DBT molecules using confocal laser scanning (Fig.~\ref{fig:DBT_molecule}b). Each bright spot represents a single dye molecule.  We place a single molecule in the focus and acquire photon stream data, which already contains a wealth of data in linear spectroscopy. Displaying the fluorescence intensity as a function of time (Fig.~\ref{fig:DBT_molecule}c) shows blinking and single-step photobleaching, indicating that the emission is indeed from a single emitter. Displayed as a histogram over the delay to the previous laser pulse (Fig.~\ref{fig:DBT_molecule}d), we can determine the fluorescence lifetime to be $\tau_{L} = 4.67$~ns, in agreement with the literature \cite{Erker2022}. Using our two photodetectors, we can obtain the cross-correlation $g^{(2)} (\tau)$ on short time scales (Fig.~\ref{fig:DBT_molecule}e). It shows an anti-bunching dip ($g^{(2)} (0) < 0.5$), another indication of a single quantum emitter. The cross-correlation is well described by a model that includes the fluorescence lifetime and  background  photons \cite{Sykora2007}.

\textbf{Two-dimensional spectroscopy} Let us now turn to nonlinear spectroscopy of single DBT molecules. We use an average power of 77~\textmu W, corresponding to a pulse energy of 136~mJ/cm$^2$ in the focus of the objective. These values are four times higher when all four fields interfere constructively. The spots in the confocal image Fig.~\ref{fig:DBT_molecule}b differ in brightness due to different orientations of the transition dipole moment relative to the linear laser polarization. We select one of the brighter molecules (marked in Fig.~\ref{fig:DBT_molecule}b) and collect about $15 \cdot 10^3$ fluorescence photons per second.

We scan the interpulse delays $t_{21}$ and $t_{43}$ over seven values between 5~fs and 62~fs, keeping the population time constant at $t_{32} = 50$~fs. Demodulating the photon stream for a linear absorption process, we find modulation amplitudes $|z| \approx 0.45$ for the delays $t_{ij}$ closest to zero. This is close to the maximum value of 0.5 when two fields interfere in the presence of two other fields that are phase-averaged. Thus, all four fields overlap well in the focus of the objective.

The maximum nonlinear modulation for the rephasing or non-rephasing process is $|z| \approx 0.03$, i.e. about 7\% of the linear effect. This means that in about 7\% of all cases where the pump pulse (formed by the first two fields) has excited the molecule, the probe pulse (fields 3 and 4) also interacts with the molecule and modifies the emission. Neglecting excited state absorption, we conclude that each pulse excites the molecule with a probability of about 7\% under these conditions. Comparing the average photon count rate of about 15~000 counts per second with the laser repetition rate (80 MHz), we estimate our photon detection efficiency to be about  0.6\%,  taking into account the quantum efficiency of the DBT (about 25\%, \cite{Erker2022}). This is a comparatively low value for single molecule spectroscopy, which is dominated by non-ideal spectral filtering of the dye emission and weak performance of the objective and photodetector in the near infrared.

We zero pad the $7 \times 7$ pixel time-domain data and Fourier transform $z(t_{21}, t_{43})$ to $Z(\omega_{21}, \omega_{43})$. 
Photon shot noise is the limiting noise source in our experiment. It leads to an expected standard deviation for each demodulation channel of 
\begin{equation}
    \sigma_{SN} = n_{pix} \sqrt{\frac{2}{ N_{total}}}
\end{equation}
where $n_{pix}$ is the number of pixels before zero padding (here 49) and $N_{total}$ is the total number of photons, summed over all time-domain pixels (see Supplementary Material, section \ref{sm:lockin}).
Depending on the phase relation used for demodulation, we get the non-rephasing (NR) and rephasing (R) spectrum (Fig.~\ref{fig:2D_spectra}). 

The purely absorptive (PA) spectrum is the average of both, and its shot noise is thus a factor of $\sqrt{2}$ smaller. 
We observe a distinct negative signal, as expected for saturating fluorescence, with a peak value of Re$(Z_{PA}) \approx - 0.47$, about 5 times the shot noise. 
To rule out other causes, such as detector nonlinearity, we acquire a laser reflection of comparable count rate and total photon number on a blank sample (Fig.~\ref{fig:2D_spectra}, 'substrate'). It shows only the noise background, not the nonlinear signal, and agrees with simulated shot noise (Fig.~\ref{fig:2D_spectra}, 'noise').

\begin{figure} 
    \includegraphics[]{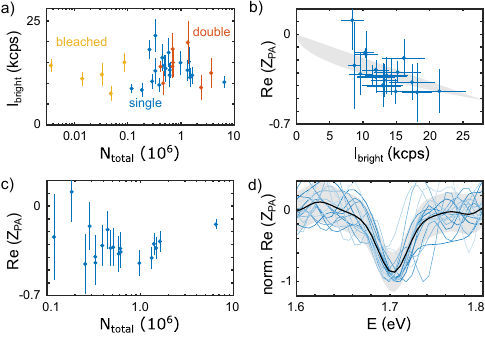}
    \caption{\textbf{Statistics of the 2D signal} a)~Correlation between the brightness of the molecule $I_{bright}$ and the total emitted photons $N_{total}$ for 32 molecules. b)~Real part of purely absorptive (PA) signal amplitude at $E = 1.705$~eV as a function of the brightness of the molecules for 19 single molecules. The gray area depicts a model assuming only orientational differences between the molecules.
    c)~Real part of PA signal amplitude as a function of the total emitted photons. d)~Diagonal cross-sections in the purely absorptive 2D signal after normalization (we excluded one outlier for this plot).}
    \label{fig:2D_statistics}
\end{figure}
 
\textbf{2D statistics} Single molecule spectroscopy provides access to distributions of spectroscopic properties. To this end, we measure the 2D spectra of more than 30 DBT molecules. Since their photostability varies (see Fig.~\ref{fig:DBT_molecule}c), we repeatedly scan the $t_{21}$ and $t_{43}$ delays until the molecule irreversibly photobleaches. Only 5 molecules bleached within the first iteration (marked yellow in Fig.~\ref{fig:2D_statistics}a). We exclude those molecules from further analysis.
Some of the time traces did not show clear single-step bleaching, possibly due to two molecules in close proximity. We also exclude these 8 molecules (red in Fig.~\ref{fig:2D_statistics}a). For all other 19 molecules (blue in Fig.~\ref{fig:2D_statistics}a), we sum all pixels at the same $(t_{21}, t_{43})$ position and calculate the 2D spectra, similar to Fig.~\ref{fig:2D_spectra}.

We excite the molecules with linearly polarized light. The excitation rate depends on the orientation of the transition dipole moment with respect to the linear polarization direction. The detection is polarization independent. The fluorescence count rate or brightness of the molecule $I_{bright}$ is proportional to the product of the excitation and detection efficiencies. The purely absorbtive signal $Z_{PA}$ is proportional to the excitation rate and does not depend on the detection efficiency. From these dependencies, one would expect a linear relationship between $I_{bright}$ and $Z_{PA}$ for in-plane oriented molecules. For out-of-plane orientations, the limiting case is a square root dependence (see Supplementary Material, section \ref{sm:orientation}).  Comparing the real part of $Z_{PA}$ at $E_{21} = E_{43} = 1.705$~eV with $I_{bright}$ for 19 single DBT molecules (Fig.~\ref{fig:2D_statistics}b), we find that the model can explain most of the variation in signal amplitude. The total number of photons detected $N_{total}$ has no visible influence on the  nonlinear signal amplitude $Z_{PA}$  (Fig.~\ref{fig:2D_statistics}c). Only the noise in the  signal decreases with increasing photon number, proportional to $1/\sqrt{N_{total}}$. About $5 \cdot 10^5$ photons are sufficient to obtain a useful  nonlinear  signal.

To compare the shape of the 2D spectra, we take a diagonal cross-section of the normalized purely absorptive signal of 19 DBT molecules. This cross-section corresponds to the linear excitation spectrum. We do not observe any significant difference (Fig.~\ref{fig:2D_statistics}d). The accessible spectral range is too small to contain relevant spectral information since our laser spectrum only partially overlaps with the absorption spectrum of the molecules. This would be different if the absorption spectrum were shifted further into the red. 

\textbf{Conclusion} Here, we presented our approach to nonlinear 2D spectroscopy of single molecules at room temperature. In more than 50\% of all attempts, we were able to measure the 2D spectra of single DBT molecules. The entire measurement took only about 5 hours of lab time. A few hundred thousand photons are sufficient to obtain meaningful 2D spectra. In this experiment, we only analyzed the photons emitted by the molecule when the stage is not in motion, so $N_{total}$ is only about $1/3$ of the total available photons. We can easily improve this by using a fast shutter to block the excitation beam while the stage is moving.

This approach has two major advantages: First, it is a Fourier technique. All $N_{total}$ detected photons contribute to the signal in each frequency domain data point. The signal-to-noise ratio scales with $\sqrt{N_{total}}$, not with $\sqrt{N}$, the number of photons per pixel. Second, rapid phase cycling in combination with photon-stream recording captures all possible interaction paths in Liouville space in a single measurement. The limited number of emitted photons is efficiently used to measure all possible four-pulse interactions at once. 

The technique will find applications ranging from plasmonic strong coupling \cite{Timmer2023} and ultrafast exciton dynamics \cite{Bakulin2016,May2020} to the role of coherence in biology \cite{Cao2020} and  light harvesting \cite{Karki2019, Tiwari2018}. Optimized sampling schemes \cite{Roeding2017,Wang2020,Bolzonello2024} and nonlinear spectroscopy via plasmonic nanostructures \cite{Schorner2020} promise to extend the parameter range even further.

\begin{acknowledgments}
We thank M.~Theisen, M.~Heindl, and C.~Schnupfhagn for their work on the pulse shaper,  F.~Paul and L.~Ludwig for their help in the early stages of this experiment, L.~Günther and S.~Goppelt for measuring the DBT emission spectra and R.~Weiner for the electronics. We acknowledge the financial support of the German Science Foundation (DFG) via IRTG OPTEXC and project 524294906. 

\end{acknowledgments}

\bibliographystyle{achemso}
\bibliography{references}
\bibstyle{achemso}
\bibdata{references}
 
\onecolumngrid
\newpage 
 \begin{center} \Large
    Supplementary material
\end{center} 

\section{Details of the setup}
\label{sm:setup}

We use a Ti:Sa laser oscillator from laser quantum (Venteon power). The p-polarised output goes to an SLM-based pulse shaper (SLM-S640, Jenoptik) in a double-pass configuration. After the pulse shaper, the polarisation state of the laser is vertical to the optical table (s-polarised). The output of the pulse shaper goes through a 3:1 beam expander, and a small pinhole sitting at the Fourier plane of the first lens of the beam expander spatially filters the Gaussian beam. The beam then enters the 4-arm cascaded interferometer (see Fig. \ref{fig:setup}). All four arms of the interferometer are identical. We use AOMs from G\&H (I-M080-2C10B11-4-GH95), beamsplitters from Newport (10RQ00UB.2), stages from Smaract (2 $\times$ SLC-2430, 1 $\times$ SLC-24150) and concave mirrors from Eksma optics (091-0125FR-200). To mount the mirrors and beamsplitters, we use Polaris mounts from Thorlabs. We build the interferometer on a custom design baseplate (designed in autodesk inventor software) and surround it with acoustic foam (Basotect) to shield it from ambient vibrations. 

After the interferometer, one output goes to the reference setup, where a reflective grating spectrally separates the beam, and a 3 cm lens focuses this spectrum on a fiber coupler. A fiber-coupled diode (OE-200-S, femto) collects the reference signal, and the output of the diode goes to an FPGA (USB-7856, National Instruments). Three phase-locked loops (PLLs) in the FPGA lock the phase difference and send trigger pulses to a time tagger (Time Tagger 20, Swabian Instruments). The other output of the interferometer goes through a 1:1 beam expander and a small pinhole, which acts as a spatial filter. This output, after reflection from a dichroic mirror (735 nm, Edmund optics), goes to home-build inverted microscope, consisting of a high NA (1.3) oil immersion objective (UPlanFl 100x/1.30, Olympus) and piezo stage (P173CD with controller E-710.3CD,  Physik Instrumente). We tune the dichroic's angle to shift the edge of the reflection spectra towards the red wavelength. The fluorescence signal is collected by the same objective, transmitted by the dichroic, and split into two parts by a beam splitter. Two identical single photon counting detectors (SPCM-AQRH series, Excelitas) collect the fluorescence signal. Two lenses and a pinhole sitting at the detection path enable confocal detection. A short pass filter (750SP) and a long pass filter (760LP) from AHF are used in the excitation and detection path to suppress the laser beam. The output signal of the SAPDs is recorded with the time tagger along with the laser trigger pulses. The time tagger data is stored on the Lab PC as photon-stream data, giving us the precise arrival times and channels that were triggered. This enables us to perform analysis either live during the measurement or later.

\section{Sample preparation}
\label{sm:sample}

We use separate solutions of DBT and PMMA in chlorobenzene. We dissolve a small amount of dibenzoterrylene (NewChem) in chlorobenzene (Carl Roth). This higher concentration stock solution is successively diluted to a concentration of about $1~\textrm{nM}$. We then combine the low concentration solution with a $1\%\textrm{wt}$ solution of polymethyl methacrylate (Sigma Aldrich, average $\textrm{M}_{\textrm{W}}=350000$) in chlorobenzene.

We then spin-coat the resulting mixture onto a quartz microscope slide (SPI Supplies, thickness $0.15-0.18~\textrm{mm}$) at $22$ rotations per second. Beforehand, we cleaned the slide in an ultrasonic bath with deionized water, acetone, and isopropanol. The spin coating results in an approximately $100 ~\textrm{nm}$ thick layer of PMMA in which individual DBT molecules are dispersed. We vary the exact dilution of the DBT solution to obtain an optimal density of dye molecules in the matrix.

\section{Modulation amplitude via lock-in detection}
\label{sm:lockin}

The AOMs mark the field of each arm by shifting its frequency by a frequency $\Omega_i$. The frequencies $\Omega_i$ are all chosen around 78 MHz so that their differences $\omega^{(c)} = \Omega_i - \Omega_j$ are in the range of a few 10~kHz (with $c = (ij) = 21, 32, 43$). Acoustic vibrations of the interferometer lead to additional frequency fluctuations of a few 100~Hz, which we detect with the reference diode. A phase-locked loop (PLL) recovers the instantaneous phase of the phase modulation across each delay.

The central idea of rapid phase cycling is that paths in Liouville space can be distinguished by the different dependence of the resulting fluorescence photon count rate on the modulation frequencies of the field. Since the absolute phase cannot be measured, the three frequency differences $\omega_{ij}$ are sufficient to reconstruct all possible frequency differences as a weighted sum
\begin{equation}
    \omega_{signal} = \sum_c m_c \, \omega^{(c)}
\end{equation}
with integer weights $m_c$. The photon count rate is periodically modulated with the frequency $\omega_{signal}$.
It is convenient to visualize the photon stream as a histogram $h$ over the instantaneous phase $\phi(t) = \omega_{signal} t$. 
\begin{equation}
    h_{signal}(\phi) = \frac{N}{2 \pi} \left[ 1 + |z| \cos \left( \phi(t) + \phi_{signal}^{(0)} \right) \right]
 = \frac{N}{2 \pi} \left[ 1 + \Re \left\{ z \, e^{i \phi(t)} \right\} \right]
\end{equation}
where $N$ is the total number of photons and $ z = |z| \exp i \phi_{signal}^{(0)}$ with $|z| \le 1$ being the complex modulation index. This is the quantity we measure by phase-resolved single-photon counting, similar to lock-in detection:
\begin{equation}
    z = \frac{2}{N} \, \sum_{j = 1}^n h_j \left( \cos \phi^{{h}}_j - i \sin \phi^{{h}}_j \right)
    \label{eq:supmat_a_binned}
\end{equation}
where $\phi^{{h}}_j$ is the phase corresponding to the center of the bin $j$ containing $h_j$ photons, and $n$ is the number of bins in the histogram, so that
\begin{equation}
    N = \sum_{j = 1}^n h_j \quad .
\end{equation}

\vspace*{\baselineskip}

In the experiment, the reference phase is provided by the three phase-locked loops related to the three interferometer delays $c$ ($c = 21, 32, 43$). The zero crossings of each phase-locked loop are recorded at times $T_j^{(c)}$. Thus, we know the reference phase $\Phi^{(c)}(t)$ at these times 
\begin{equation}
    \Phi^{(c)}(T_j^{(c)}) = j \, 2\pi - \Phi^{(c)}_0
\end{equation}
and linearly interpolate for the times $t$ in between. The phase offset $\Phi^{(c)}_0$ results from the phase lag in the AOM frequency generation and amplification and in the PLL.

\vspace*{\baselineskip}

We measure the arrival times $t_i$ of the fluorescence photons and assign a phase $\Phi_i$ to each photon for each demodulator. 
\begin{equation}
    \Phi_i = \sum_c m_c \, \Phi^{(c)}(t_i) \quad .
\end{equation}
We compute the modulation index $z$ as 
\begin{equation}
    z = x + i y = \frac{2}{N} \sum_{i = 1}^N \cos \Phi_i - i \sin \Phi_i
\end{equation}
which is eq.  \ref{eq:supmat_a_binned} in the limit of very small bin sizes, e.g., the temporal resolution of the time tagger.
We limit the integration domain so that we integrate over full $2\pi$ periods, i.e. 
\begin{equation}
    \Phi_{N} \le \Phi_1 + m \, 2 \pi < \Phi_{N+1}
\end{equation}
with the largest possible integer $m$. Considering all photons, this is a small correction since the modulation period is about a factor of 100 faster than the pixel dwell time.

\vspace*{\baselineskip}

Let us estimate the uncertainty $\delta x$ in the real part $x$ of $z$. The uncertainty in the imaginary part $y$ will be the same.
\begin{equation}
   \delta x  = \frac{2}{N} \sqrt{  \sum_{j = 1}^n  \left( \delta h_j   \cos \Phi^{{h}}_j \right)^2 } \quad .
\end{equation}
We are primarily interested in small signals. The sinusoidal modulation in the histogram is, therefore, small compared to its mean. In the experiment, we get nonlinear modulation values of a few percent. Let us assume that all $h_j$ are equal ($h_j = N/n$) and sum to the total number of photons $N$. The uncertainties $\delta h_j$ are also equal and approximately $\sqrt{N/n}$ since the photon flux follows a Poisson distribution. We thus get
\begin{equation}
    \delta x  = \frac{2}{N}  \sqrt{\frac{N}{n}} \,  \sqrt{  \sum_{j = 1}^n  \left(  \cos \Phi^{{h}}_j \right)^2 } \quad .
\end{equation}
 The sum gives $n/2$ so that 
\begin{equation}
    \delta x  =  \sqrt{\frac{2}{N}} \quad .
\end{equation}

The two-dimensional discrete Fourier transform of $z(t_1, t_2)$ to $Z(\omega_1, \omega_2)$ is a weighted sum over the components $n \times n$ of $z$ where the weights are phase factors with an absolute value of one:
\begin{equation}
    Z_{p,q} = \sum_{j,k=0}^{n-1} z_{j,k} \, e^{- 2 \pi i j p /n}  \, e^{- 2 \pi i k q /n} 
\end{equation}
The uncertainty $\delta X$ in the real part of $Z$ is
\begin{eqnarray}
    \delta X =  & \sqrt{ \sum_{j,k=0}^{n-1} | \delta x \, e^{- 2 \pi i j p /n}  \, e^{- 2 \pi i k q /n} |^2 } \\
             =  & \sqrt{ \sum_{j,k=0}^{n-1} | \delta x  |^2 } \\
             =  & \sqrt{ n^2 | \delta x  |^2 } \\
    = & n^2  \sqrt{\frac{2}{N_{total} }}  
\end{eqnarray}
with $N_{total} = n^2 N$. Zero padding does not contribute uncertainty, so it does not change $\delta X$ as long as we define $n^2$ as the number of non-zero pixels in the time domain.

\section{Excitation and detection probability as a function of transition dipole orientation}
\label{sm:orientation}

We assume that excitation and fluorescence emission occur via the same transition dipole. Its orientation is defined by an in-plane angle $\xi$ and an out-of-plane angle $\theta$. When excited with linear polarized light, the excitation probability $\eta_{ex}$ is 
\begin{equation}
    \eta_{ex} \propto \cos^2 \theta \, \cos^2 \xi \quad .
\end{equation}
When the detection path is polarization independent, the probability of detection is
\begin{equation}
    \eta_{det} \propto \cos^2 \theta \, \left(  \cos^2 \xi + \sin^2 \xi \right) = \cos^2 \theta \quad .
\end{equation}
The brightness of a molecule is proportional to the product of the two, and the nonlinear modulation amplitude is proportional only to the excitation, i.e.
\begin{align}
    I_{bright} &  = I_0 \, \eta_{ex} \, \eta_{det} = I_0 \, \cos^4 \theta \, \cos^2 \xi \\
    Z_{mod}^{NL} & = Z_0 \, \eta_{ex} =  Z_0 \, \cos^2 \theta \, \cos^2 \xi
\end{align}
so that
\begin{equation}
    \frac{ Z_{mod}^{NL}}{Z_0}  = \frac{1}{\cos^2 \theta} \, \frac{I_{bright}}{I_0} \quad .
\end{equation}
If all molecules are oriented parallel to the glass surface, then $\theta = 0$, and we get a simple proportionality. 

Otherwise, the variation of $\theta$ results in a range of possible modulation amplitudes for a given brightness. The lower limit is given by the highest value, i.e., $\cos^2 \theta = 1$. The lowest value of $\cos^2 \theta$ is reached when $\cos^2 \xi = 1$, i.e. when $\cos^2 \theta = I_{bright} / I_0$, where $I_0$ is the maximum brightness of the molecule and $Z_0$ the maximum nonlinear modulation. The limits are thus given by
\begin{equation}
    \frac{I_{bright}}{I_0} \le    \frac{ Z_{mod}^{NL}}{Z_0}  \le  \sqrt{\frac{I_{bright}}{I_0} } \quad .
\end{equation}
Numerical simulations show that a random orientation of the molecules leads to an almost constant probability density between these limiting cases.

\end{document}